\documentclass[12pt]{article}%
\pdfoutput=1
\usepackage{cite}
\usepackage{fix-cm}
\usepackage{color}
\usepackage[usenames,dvipsnames]{xcolor}
\usepackage{heck}
\usepackage{amsmath}
\usepackage{amsfonts}
\usepackage{amssymb}
\usepackage{graphicx}
\usepackage{multicol}
\usepackage{hyperref}
\usepackage{setspace}
\usepackage[all]{xy}
\usepackage{verbatim}
\usepackage{color}
\usepackage{epsfig}
\usepackage[margin=1in]{geometry}

\providecommand{\U}[1]{\protect\rule{.1in}{.1in}}
\def\be{\begin{equation}}
\def\ee{\end{equation}}
\def\bea{\begin{eqnarray}}
\def\eea{\end{eqnarray}}
\def\spc{\hspace{1pt}}

\newcommand{\smpc}{\hspace{.5pt}}
\def\is{ \! &  \! = \!  & \!   }
\def\ra{\rangle}
\def\la{\langle}
\def\nspc{\hspace{-.5pt}}

\definecolor{gray}{rgb}{0.5,0.5,0.5}
\begin{document}
\def\iimath{\dot{\imath}}
\addtolength{\abovedisplayskip}{.2mm}
\addtolength{\belowdisplayskip}{.2mm}
\addtolength{\baselineskip}{.3mm}
\addtolength{\parskip}{.3mm}
\renewcommand{\footnotesize}{\small} 
\def\dd{d} %{\raisebox{-.5pt}{d}}
\def\li{{\bigl |}}
\def\ra{{\bigr\rangle}}
\def\la{{\bigl\langle}}

\date{August 2013}

\title{Bekenstein-Hawking Entropy \\[3mm]
as Topological Entanglement Entropy}

\institution{PU}{\centerline{${}^{1}$Department of Physics, Princeton University, Princeton, NJ 08544, USA}}

\institution{PCTS}{\centerline{${}^{2}$Princeton Center for Theoretical Science, Princeton University, Princeton, NJ 08544, USA}}

\authors{Lauren McGough\worksat{\PU}\footnote{e-mail: {\tt mcgough@princeton.edu}} and Herman Verlinde \worksat{\PU}\worksat{\PCTS}\footnote{e-mail: {\tt verlinde@princeton.edu}}}

\abstract{Black holes in 2+1 dimensions enjoy long range topological interactions similar
 to those of non-abelian anyon excitations in a topologically ordered medium. Using this observation, we compute the topological entanglement entropy of BTZ black holes via the established formula $S_{\rm top}  = \log(S^{\spc a}_0)$, with $S_b^{\spc a}$ the modular S-matrix of the Virasoro characters $\chi_a(\tau)$. We find a precise match with the Bekenstein-Hawking entropy. This result adds a new twist to the relationship between quantum entanglement and the interior geometry of black holes. We generalize our result to higher spin black holes, and again find a detailed match. We comment on a possible alternative interpretation of our result in terms of boundary entropy.}

\maketitle

\noindent
{\large \bf 1.  Introduction}

\medskip

The close relation between black hole physics and thermodynamics provides crucial guidance to
the search for consistent quantum theories that incorporate gravity. In particular, it indicates that pure quantum gravity -- i.e. any attempt to directly quantize the Einstein lagrangian, without the addition of any matter degrees of freedom -- is unlikely to give rise to a complete theory. Metric excitations alone seem insufficient to account for the microscopic entropy of black holes, quantified via the Bekenstein-Hawking formula \cite{BH}
\bea
\label{sbh}
S_{\rm BH} =   \frac {\rm Area} {4G_N}\, .
\eea
A more promising perspective is that general relativity represents a long range effective theory with dynamical rules that encode the quantum information flow of underlying elementary degrees of freedom.
This point of view is supported by string theory realizations of black hole space-times,
in which the B-H formula (\ref{sbh}) has been successfully matched with the microscopic entropy
of the constituent strings, D-branes and their excitations \cite{SV}.

Another powerful diagnostic tool
is the geometric entanglement entropy \cite{GEE}, which has received much recent attention. Let $A$ denote a 
region of space, such as the interior of a black hole, and $B$ its complement, all of space outside of $A$.
The density matrix associated with $A$ is $\rho_A = \tr_B\bigl( \li \psi\ra \la \psi \li\bigr)$, where $\li \psi\ra$ is typically taken to be the ground state of the system, and the trace is over all states of $B$. The von Neumann entropy 
$$
S_A = - \tr(\rho_A \log \rho_A)\,
$$
quantifies the total entanglement between region $A$ and its complement $B$. 

The importance of  entanglement for the microscopic structure of space-time is only beginning to emerge. There are tantalizing  hints of a deep connection, most notably the Ryu- Takayanagi formula 
\cite{RT}\cite{van} and the firewall debate \cite{AMPS}. In this note, we study this relationship in 2+1-D AdS space-times. Einstein gravity in 2+1 dimensions  has special characteristics, akin to Chern-Simons (CS) gauge theories  \cite{WG,WCS} that capture the infrared properties of quantum critical systems with topological order \cite{KP,LW}. Massive spinning point particles and black holes enjoy long range interactions that generalize the braiding relations of particles with non-abelian statistics \cite{nayak}. In addition,  the system possesses a ground state degeneracy that is sensitive to the global space-time topology. In condensed matter systems, such as those exhibiting the fractional quantum Hall effect, these remarkable properties emerge because the ground state of the underlying medium is deeply entangled
\cite{fradkin}. Quantum gravity in 2+1 dimensions should be thought of in the same way:
as the effective theory that captures the topological Berry phases of the ground state wave function. It is through these topological interactions that the quantum order of the microphysical medium manifests itself. 

Topological entanglement entropy provides a quantitative measure of this long range quantum order\cite{KP,LW}. Consider a region $A$ with disk-like topology and a smooth boundary of length $L$. In a gapped quantum many-body system, the geometric entanglement entropy of $A$ has the form
\bea
\label{sent}
S_A \is \alpha\spc {L} + S_{\rm top} + \ldots
\eea
where $...$ indicate terms that vanish in the limit  $L \to \infty$. The first term arises from short wavelength modes straddling the boundary of the entangling region. The pre-coefficient $\alpha$ is non-universal, and  depends on the UV cut-off. The constant term $S_{\rm top}$ is the topological entanglement entropy; it represents a universal characteristic of the many-body vacuum state \cite{KP,LW}. In the above sign convention, it is typically $\leq 0$. 
It can be isolated from the length term by dividing the region $A$ into three or more segments and taking a suitable linear combination of the resulting entanglement entropies
in which the boundary terms cancel. Since topological entanglement entropy survives in the long distance limit, $L \to \infty$, it can be calculated by means of the low energy topological field theory
that describes the braiding properties of the quasi-particle excitations. In case the region $A$ contains a single excitation labeled by some charge $a$, one finds that \cite{KP,LW, fradkin}
\bea
\label{stop}
S_{\rm top} = \log \bigl(d_a/ {\cal D}\bigr) = \log( S_0^{\spc a}).
\eea
Here ${\cal D}$ and $d_a$ are the quantum dimensions of the medium and the $a$ excitation, respectively.
 $S_0^{\spc a}$ denotes a matrix element  
of the modular $S$-matrix of the 1+1-dimensional CFT that describes the edge excitations of the topologically ordered medium. The quantity $S_{\rm top}$ has the key property that it does not depend
on the size or geometry of the region $A$.

Topological entanglement entropy and black hole entropy seem unrelated.
The B-H formula of 2+1-D black holes \cite{BTZ} relates the entropy to the length of the event horizon via
\bea
\label{sbhl}
S_{\rm BH} \is \frac  {{\rm Length}(\Gamma)} {4G_N}\, .
\eea 
This looks similar to the non-universal length term in  (\ref{sent}), except that the coefficient $\alpha$ is now a universal constant. Because of this similarity,
many authors have suggested that the B-H formula may also have an interpretation as geometric entanglement entropy \cite{GEE}. There is growing evidence that this is indeed the case \cite{RT,van,VV}. This is an important insight. In particular, it indicates that black holes are typically in a near-maximally entangled state.

However, there is one unsatisfactory aspect to relating the length contributions in (\ref{sbhl}) and (\ref{sent}). Unlike the first term in (\ref{sent}), the B-H formula (\ref{sbhl}) is universal and robust. In this respect, black hole
entropy seems more similar to the universal constant contribution in (\ref{sent}).  
Could it be that the 2+1-D black hole entropy (\ref{sbhl}) can be identified with the universal topological entanglement entropy associated with the black hole space-time?

\begin{figure}[t]
\begin{center}
\vspace{-.7cm}
\includegraphics[scale=.42]{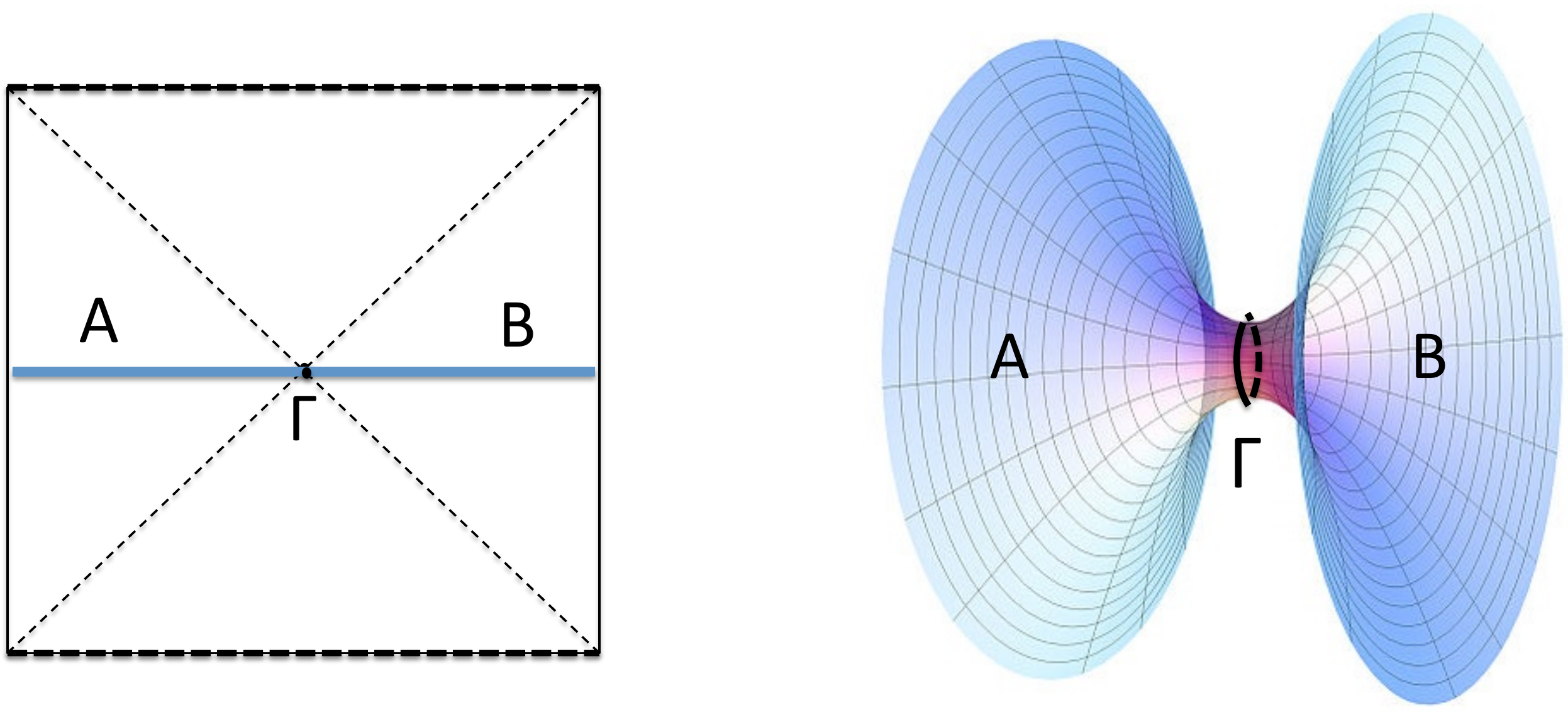}
\caption{\small  The black hole horizon forms a geodesic $\Gamma$.  The entanglement entropy between the inside and outside regions $A$ and $B$ is equal to Length$(\Gamma)/4$.}
\end{center}
\vspace{-0.5cm}
\end{figure} 

At a first glance, this seems implausible: the B-H formula does not appear topological, for it is proportional to a length.
How, then, could this be true? Fig.~1 shows a Penrose diagram of an eternal BTZ black hole of mass $M$ (and spin $J\!=\!0$) and a spatial slice with an
Einstein-Rosen bridge connecting the two sides. The horizon is a geodesic: it has
minimal length for the given topology of $\Gamma$. So we can view
Length$(\Gamma)$ as a common property of all loops with the same topology of $\Gamma$. 
In other words, Length$(\Gamma)$ should not be viewed as a geometric property of a loop, but as a quantum number of the black hole state, determined by its mass $M$ and spin~$J$. 

Let us now view the black hole as a localized defect of a topological ordered system, and treat  $M$ and $J$ in the same way as the charge label $a$ in (\ref{stop}). This interpretation is natural given that
2+1-D gravity can be written as a ${\cal G} = SL(2,\mathbb{R}) \times SL(2,\mathbb{R})$ Chern-Simons theory \cite{WG},
in which the black hole state represents a heavy particle with a large ${\cal G}$ charge.
The edge states of  2+1-D gravity are described by Liouville theory \cite{HV, carlip, ZZ,TT1}, the universal conformal field
theory associated with the Virasoro algebra. Although Liouville theory is a non-rational CFT with central charge $c = \frac 3 2 \ell \gg 1$, it shares many features with rational CFTs \cite{ZZ,TT1}. In particular, the Virasoro conformal blocks form a unitary representation of the modular and braid group, characterized by the quantum group  ${\cal U}_q\bigl({\mathfrak{sl}(2,\mathbb{R})}\nspc \times {\mathfrak{sl}(2,\mathbb{R})}\bigr)$. This representation is infinite-dimensional, and
modular and fusion relations are expressed as integrals rather than finite sums. Nonetheless, one can identify analogs of quantum dimensions and of the modular S-matrix $S_b^{\spc a}$. 

We  can  thus apply the same formulas (\ref{stop}) to compute the topological entanglement entropy associated with the black hole excitation.
Using the proper identification of a black hole of mass $M$ and $J$ with a superselection label $a$ of Liouville CFT, we find a precise match
\bea
\label{magic}
\qquad \quad S_{\rm BH} \, = \, {\rm log}\bigl(S_0^{\spc a} \bigr), \qquad \qquad \mbox{\small $a = (M,J)$}.
\eea
We describe the details of this calculation in the following sections. To test the robustness of our
result, we also consider the higher spin black holes, and find an encouraging match with known results.

This identification and interpretation of the Bekenstein-Hawking entropy as topological entanglement entropy raises many conceptual questions.  
Why does the computation of the topological entanglement entropy reproduce the microscopic entropy? 
What does our computation say about the applicability and validity of pure quantum gravity in 2+1 dimensions?
What is entangled with what? What does the calculation imply for the firewall controversy?  
We address these questions in the concluding section.

\bigskip
\bigskip

\noindent
{\large \bf 2.  BTZ Black Hole}

\medskip

We briefly summarize the main properties of the BTZ black hole \cite{BTZ, carlip}. From now on we put $G_N=1$,
so  $\ell$ denotes the AdS$_3$ curvature radius in Planck units.

AdS$_3$ can be identified with the universal covering space of the group $SL(2,\mathbb{R})$, 
and has isometry group  ${\cal G} = SL(2,\mathbb{R}) \times  SL(2,\mathbb{R}).$
The BTZ black hole space-time is obtained by taking the quotient of AdS$_3$
with a hyperbolic group element $(h_+,h_-) \in {\cal G}$, acting via 
\bea
\label{quotient}
g \sim  h_+ g h_- , \  \ & & \ \  h_\pm = e^{\mbox{\small $\pi (r_+ \pm r_-) \sigma_3/\ell$}}.
\eea
The quotient describes  a stationary and axially symmetric black hole
% The metric in Schwarzschild coordinates takes the form
%\bea
%\label{btz1}
%ds^2 \is-\frac{(r^2\! - r_+^2)(r^2\!-r_-^2)}{r^2\ell^2} \spc dt^2 + \frac{r^2\ell^2} {(r^2\! - r_+^2)(r^2\!-r_-^2)} \spc dr^2 + r^2 \Bigl(d\phi -\frac{r_+r_-}{r^2} \spc dt\Bigr)^2 
%\eea
with an outer event horizon at $r_+$ and an inner Cauchy horizon at $r_-$. 
The BTZ  metric can be written as
%$
%r_\pm^2 =\textstyle 4 G M \ell^2 \Bigl(  1 + \sqrt{1 - \bigl({J}/{M\ell\spc}\bigr)^2}\,\Bigr)
%$ 
\bea
\label{btz2}
ds^2 \is -4\ell \bigl(\Delta_+ du^2 + \Delta_- dv^2\bigr) + d\rho^2+ \bigl(\ell^2 e^{2\rho} + \Delta_+\Delta_-\bigr) du dv.
\eea
The two radii $r_\pm$ and the constants $\Delta_\pm$
are related to the black hole mass and spin  via
 \bea
 \label{mjrel}
M = \frac{ r_+^{\spc 2} + r_-^{\spc 2}}{8\ell^2}, \ \ & & \  \  J = \frac{r_+r_-}{4\ell}, \qquad 
\Delta_\pm = \, \frac{ (r_+ \pm r_-)^2}{16 \ell}  = \frac 1 2\bigl(\ell M \pm J).
\eea

Einstein gravity in 2+1 dimensions can be formulated as a CS-type gauge theory by introducing the
dreibein $e^a$ and spin connection $\omega^a$.
The linear combinations
$A^a_{\pm} = \omega^a \pm \frac 1 \ell e^a$ form two $SL(2,\mathbb{R})$ connections, in terms of which the torsion constraint and Einstein equation take the form of flatness constraints \cite{WG}. The group elements 
$h_\pm$ in Equation (\ref{quotient}) coincide with the holonomies of  $A^a_\pm$ around the black hole. 
In general, $SL(2,\mathbb{R})$ holonomies come in three types, depending on
whether the conjugacy class of the group element is hyperbolic, parabolic, or elliptic. 
For a black hole, both holonomies are in a hyperbolic conjugacy class.

%They are the transition function acting on the wave function of a test particle that traverses a closed path around the black hole.

The Bekenstein-Hawking entropy of the BTZ black hole is equal to 
\bea
\label{sbtz}
S_{\rm BH} = \frac{2\pi r_+}{4}. %=\textstyle 2\pi M \ell^2 \Bigl(  1 + \sqrt{1 - \bigl({J}/{M\ell\spc}\bigr)^2}\,\Bigr).
\eea
This formula has been reproduced in numerous dual CFT realizations of string theory on AdS$_3$ by counting the number of states at energy $M$ and with angular momentum $J$. 
 Below we will give a new derivation and interpretation.

\begin{figure}[t]
\begin{center}
\vspace{-1,2cm}
\includegraphics[scale=.54]{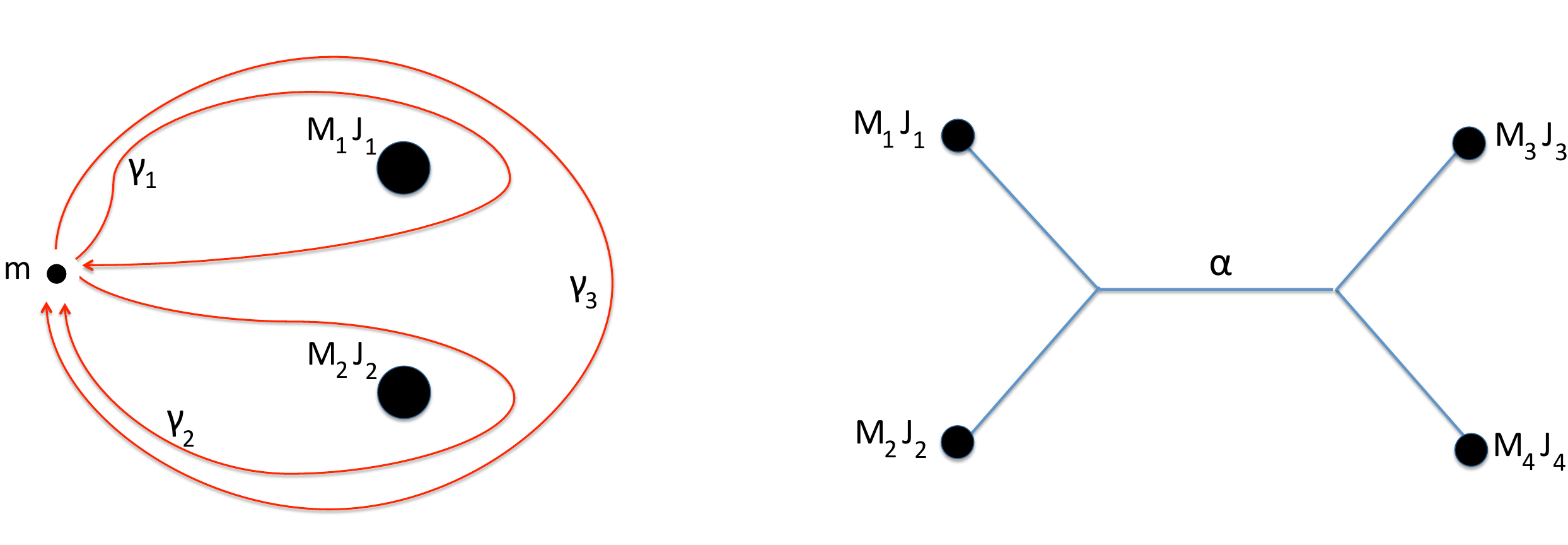}
\caption{\small The classical space-time geometry is specified by the holonomies around the paths $\gamma_i$. In the quantum theory, states are identified with conformal blocks of 2D Liouville CFT. }
\end{center}
\vspace{-0.5cm}
\end{figure} 

\bigskip
\bigskip

\noindent
{\large \bf 3. Quantum Geometry}

\medskip

Quantum geometry arises from quantizing the phase space of space-time geometries.
As an example, Fig.~2 indicates the geometry of two BTZ black holes,
specified by the $SL(2,\mathbb{R}) \times SL(2,\mathbb{R})$ holonomies around the paths $\gamma_i$. 
These holonomies are determined, up to overall conjugation, by the mass, spin, center of mass energy and total angular momentum of the two black holes. This description generalizes to  any number of point particles and black holes \cite{WG}.
The space of $SL(2,\mathbb{R})$ holonomies is isomorphic to Teichmuller space, the space of constant negative curvature metrics on a 2-D surface. 
%Teichmuller space comes with a natural 
%Kahler form $\omega$, and thus looks like a phase space. 
The phase space of 2+1-D Einstein gravity consists of two copies of Teichmuller space \cite{HV}.

  The problem of quantizing Teichmuller space has been solved \cite{ZZ,TT1,TT2}.  It  gives rise to a Hilbert space of states that can be identified with
the linear space spanned by the chiral conformal blocks of 2-D Liouville theory \cite{ZZ,TT1,TT2}
\bea
S_L(\varphi) \is \frac 1 {4\pi} \int\!\! d^2\xi\, \Bigl[\spc \frac 1 2 (\partial\smpc \varphi)^2 + Q R\varphi + \mu e^{b \smpc \varphi}\Bigr]\, , \quad 
 \quad Q = b + b^{-1}.
\eea 
This correspondence generalizes the well-known relationship between Chern-Simons theories 
and WZW conformal field theory \cite{WCS}. The dictionary is analogous. The 2-D CFT describes the massless edge excitations at the boundary of the AdS space, and supports a unitary
representation of the asymptotic symmetry group of the bulk theory. For
pure AdS$_3$ gravity, this symmetry group takes the Virasoro algebra with central charge \cite{BRH}
\be
c = 1 + 6 Q^2 = {3\ell}/{2} \, .
\ee

States of 2+1-D gravity with particle and black hole excitations in the bulk are identified with the product of left and right conformal blocks of Liouville CFT with corresponding vertex operator insertions. They
enjoy a $q$-deformed version of the monodromy properties of the classical geometry, governed by the non-compact quantum group ${\cal U}_q\bigl({\mathfrak{sl}(2,\mathbb{R})}\nspc \times {\mathfrak{sl}(2,\mathbb{R})}\bigr)$ with  $q = \exp({i\pi b^2})$.
Liouville vertex operators take the general form  $V_\alpha =  e^{\alpha_+ \varphi_+} \spc e^{ \alpha_-\varphi_-},$ and are in one-to-one correspondence with unitary highest weight representations
 of the left and right Virasoro algebra with conformal weights $\Delta_\pm = \alpha_\pm (Q-\alpha_\pm)$.
 The physical range of positive conformal weights splits into two separate regimes of  Liouville momenta
\bea
\alpha_\pm \in [\spc 0,  \textstyle \frac 1 2 Q \spc ] \, \cup\, \bigl( \frac 1 2 Q   + i\smpc \mathbb{R}^+\bigl).
\eea

The Liouville equation prescribes that the metric have constant negative curvature everywhere except at the location of the vertex operators. 
Vertex operators with real Liouville momentum in the interval $[0, \frac 1 2 Q]$ create elliptic solutions, which are local cusps specified by a patching function in the elliptic conjugacy class of the isometry group ${\cal G}$. Vertex operators with complex momenta of the form $\frac 1 2 Q   + i\smpc \mathbb{R}^+$ create hyperbolic solutions,  which are macroscopic holes in 2-D space identified with the spatial section of BTZ black hole geometries (as shown in Fig.~1 and Fig.~3.). We may parametrize the Liouville momenta in this range as 
\bea
\label{dict}
\alpha_\pm = \textstyle \frac 1 2 Q  + i p_\pm, \qquad \qquad
\Delta_\pm  =  p_\pm^2 + \textstyle \frac 1 4  Q^2 .
\eea
These relations, combined with Equations (\ref{btz2}) -(\ref{mjrel}), specify a precise dictionary between the classical data of the BTZ black hole and the quantum data of Liouville theory.  For later reference, we make note that in the semiclassical regime $p_\pm \gg b \gg 1$, the relations between the Liouville momenta $p_\pm$ and the conjugacy class of the holonomies $h_\pm$ in (\ref{quotient}) simplify to
\bea
\label{semrel}
r_\pm = \spc 4 b (p_+ \pm p_-), \qquad \qquad b^2 \spc = \spc   \ell/4\, .
\eea

\begin{figure}[t]
\begin{center}
\vspace{-.8cm}

\includegraphics[scale=.37]{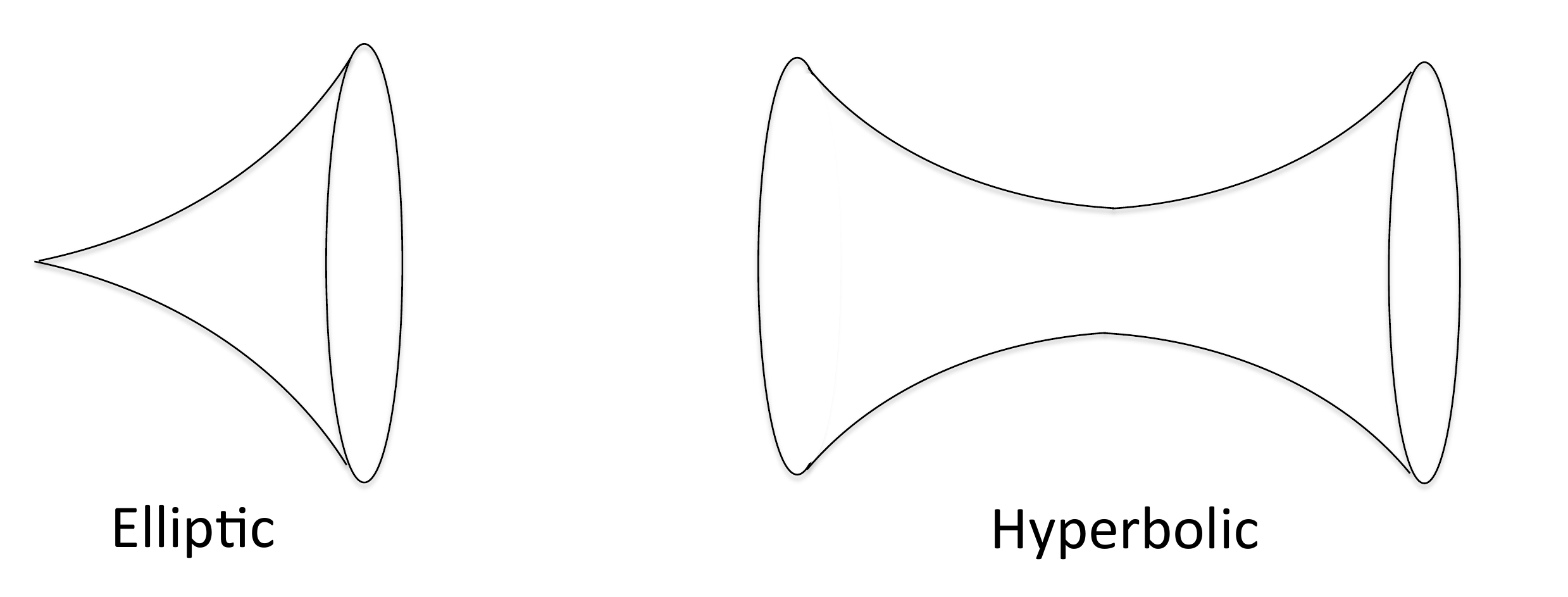}
\caption{\small Liouville vertex operators fall into two classes. Those with $\Delta< \frac 1 4 Q^2$  create {\it elliptic}  solutions (punctures), those with $\Delta > \frac 1 4 Q^2$ create {\it hyperbolic} solutions (macroscopic holes) \cite{Nati}.
 }
\end{center}
\vspace{-0.5cm}
\end{figure}

Most of the above dictionary was known before the discovery of gauge/gravity duality.
An important insight from AdS/CFT is that the bulk theory can not be pure gravity. Gravity in 2+1 dimensions describes how massive localized excitations interact at long distances, but it does not specify the hyperfine structure of the excitation spectrum of the bulk string theory. 

The situation in the 1+1-D boundary theory is analogous. 
Liouville theory has a continuous spectrum of conformal dimensions, and is therefore
capable of describing any set of Virasoro representations. However, it does not prescribe 
the spectrum of some given unitary CFT. Liouville theory is similar to a 
non-compact space with a continuous spectrum of wave solutions; choosing a specific
 CFT realization of AdS$_3$ is like putting the wave solutions in a finite box, so that the spectrum
 becomes discrete and countable.

\bigskip
\bigskip
\medskip

\noindent
{\large \bf 4. Quantum Dimension}

\medskip

An important ingredient of our story is the {\it quantum dimension} associated with a local excitation
in a topological quantum field theory. 
We first recall the definition and properties of  the quantum dimension of a topological QFT associated to a rational CFT. We then generalize 
to the case of interest, the non-rational $c > 25$ Virasoro CFTs.

The most physical definition of the quantum dimension is as follows. Let ${\cal H}_a(N)$ denote the Hilbert space  of the 2+1-dimensional topological QFT spanned by all states that contain $N$  local excitations of charge $a$. It can be shown that the dimension of this Hilbert space grows exponentially at large $N$ according to
\bea
\label{count}
{\rm dim}\smpc  {\cal H}_a(N) \propto \bigl(d_a\bigr)^N.
\eea
The number $d_a$ defines the quantum dimension of the excitation $a$.

Quantum dimensions are linked with the fusion algebra \cite{EV}. 
A local excitation with charge $a$ corresponds to a
primary vertex operator $V_a$ in the CFT. The operator product of $V_a$ and $V_b$
can be expanded as a sum of operators $V_c$. 
For rational CFTs, the fusion coefficients $N_{ab}{}^c$ are integers
that specify the multiplicity of $V_c$ in this expansion.  
The fusion algebra is commutative and associative, and admits a one-dimensional representation 
 $d_a d_b = \sum_c N_{ab}{}^c d_c$.
This relation can be used to prove the result (\ref{count}).

Quantum dimensions can be thought of as the character of the superselection sector ${\cal H}_a$
associated with the primary vertex operator $V_a$. States in ${\cal H}_a$ are obtained by acting with symmetry generators on the primary state $|a\rangle =  V_a|0\rangle$. The partition function 
\bea
\chi_a(\tau) \is \tr_{{\cal H}_a}\bigl( e^{i\pi \tau L_0}\bigr)
\eea
is called the character of the sector ${\cal H}_a$. The quantum dimension $d_a$ is obtained by taking the 
$\tau \to 0$ limit of the ratio of $\chi_a(\tau)$ with the identity character \cite{EV}
\bea
\label{dadef}
d_a \is \lim_{\tau \to 0} \frac{\chi_a (\tau)}{\chi_0(\tau)} \, .
%= \lim_{\tau \to 0} \frac{\sum_b S_a^b\spc \tilde{\chi}_b (\tau)}{\sum_c S_0^c\spc \tilde{\chi}_c(\tau)} = \frac{S_0^{\smpc a}}{S_{\smpc 0}^0}
\eea
This definition naturally explains why the quantum dimensions generate the fusion algebra.
It also allows us to re-express $d_a$ in terms of the modular $S$-matrix, which describes the
transformation properties of the characters under the modular transformation $\tau \to -1/\tau$
\bea
\label{modular}
{\chi}_a\bigl(-1/\tau\bigr) \is \sum_b S_a^{\spc b} \spc \chi_b(\tau).
\eea
Applying the modular transformation (\ref{modular})
to (\ref{dadef}), and using that for $\tau \to 0$ the dominant term in
the sum comes from the identity character, one finds that
\bea
\label{dsrel}
d_a = \frac{S_0^{\smpc a}}{S^{\smpc 0}_0}. % \qquad D 
\eea
This formula for the quantum dimension holds for rational CFTs and plays a key role in the computation of 
topological entanglement entropy. We will use this connection momentarily.

First, we need to generalize the above formulas to the case of non-rational CFTs relevant to 2+1-D gravity. The modular geometry of Liouville theory is by now quite well-developed \cite{ZZ, TT1,TT2}, and many of the RCFT formulas have found direct non-rational analogs. There are two main differences.
Since the spectrum of allowed conformal dimensions is continuous,  modular transformations and fusion coefficients are no longer described by discrete sums and finite matrices but by integrals and continuous distributions. Another important difference is that the identity representation plays a rather distinct role.
In spite of these dissimilarities, there still exists a natural analog of the notion of quantum dimension.
  
Let us follow the naive route and simply apply the formula (\ref{dsrel}). The $c>25$ Virasoro characters
for the continuous representations of conformal weight $\Delta > \frac 1 4 Q^2$ are given by
\bea
\chi_p(\tau) = \frac{e^{i\pi \tau p^2}}{\eta(\tau)}, \ \ & &  \ \  \Delta_p = p^2 +\textstyle \frac 1 4 Q^2,
\eea
where the Dedekind $\eta$-function $\eta(\tau) = q^{c/ {24}} \prod_{n>0} (1 -q^n)$ with $q\equiv e^{2\pi  i \tau}$.  The identity character 
\bea
  \chi_0(\tau) =  \frac{ e^{-i \pi \tau \frac{Q^2\!\!} 4}(1-e^{i\pi \tau})}{\eta(\tau)} , \ & & \ \Delta = 0
\eea
%exhibits the presence of the null state $L_{-1} \bigl|\spc 0\spc \bigr\rangle$. The identity character 
follows the following
modular transformation property \cite{ZZ}
\bea
\label{modtra}
{\chi}_0 \bigl(-1/\tau\bigr) %\bigl(-\frac 1 \tau\spc\bigr)\spc
\, =  \int_0^\infty\!\!\!\! dp \;  S^{\spc p}_0 \spc \chi_p(\tau)\quad\, \\[3.5mm]
\label{smeas}
\qquad S^{\smpc p}_0 \, = \, 2\sqrt{2}\sinh (2\pi b\smpc p ) \sinh (2\pi b^{-1}\nspc p)\, .
\eea
Note that i)  $S_0^{\spc p}$ is not a matrix entry of a finite matrix, but a measure on the 
continuous series of
Virasoro representations, ii) $S_0^{\spc p}$ grows exponentially with $p$, and iii)
 the identity representation itself does not appear on the right-hand side of (\ref{modtra}).

Boldly applying the formula (\ref{dsrel}), we find that, up to an irrelevant overall constant, the quantum dimension of the
representation with Liouville momentum $\alpha = \frac 1 2 Q + i p$ is given by
\bea
\label{pmeas}
d(\alpha) \is \sinh (2\pi b\smpc p ) \sinh (2\pi b^{-1}\nspc p).
\eea
 This quantity $d(\alpha)$ indeed plays a special role in Liouville modular geometry~\cite{TT2}.
As mentioned above, the representation theory and modular geometry of the Virasoro conformal blocks with 
$c > 25$ is associated with the representation theory of the quantum group ${\cal U}_q\bigl({\mathfrak{sl}(2,\mathbb{R})}\bigr)$. This quantity (\ref{pmeas}) naturally appears in this context as the weight of a representation in the Peter-Weyl or Plancherel decomposition of the space of functions on the quantum group.  This so-called Plancherel measure is the most natural counterpart of the quantum dimension in the nonrational case.

%$$
%\dd(\alpha) = \sin \bigl(2\pi b(\alpha\nspc -\nspc\delta)\bigr) \sin \bigl(2\pi b^{-1}(\alpha\nspc-\nspc\delta)\bigr )
%$$
\bigskip
\bigskip

\noindent
{\large \bf 5. Topological Entanglement Entropy}

\medskip

Gravity is  topological in the sense that every observable must be coordinate invariant. In 2+1 dimensions, this topological nature is enhanced by the fact that there are
no graviton excitations, and that the metric, outside of matter distributions, locally always looks the same.
Our proposal is that from a microscopic perspective, these properties emerge because gravity is the long distance description of the highly entangled ground state of a topologically ordered system close to a quantum critical point. For analogous
condensed matter systems, the natural diagnostic for the presence of topological order is the topological entanglement entropy introduced in \cite{KP,LW}. Let us briefly recall its definition.

%\begin{figure}[b]
%\begin{center}
%\vspace{0cm}
%\includegraphics[scale=.36]{kitpres.pdf}
%\caption{\small The definition  of topological entanglement given in \cite{kp} makes use of the division of the plane into four subregions $A,B,C$ and $D$.}
%\end{center}
%\vspace{-1cm}
%\end{figure} 

To compute the topological entanglement entropy $S_{\rm top}$ of a disk-shaped region $R$ with the
outside $D = R^c$, one first divides the interior of $R$ into three sectors $A$, $B$ and $C$.
Let $S_A = - \tr\rho_A \log \rho_A$ denote the von Neumann entropy of the density matrix $\rho_A$ associated with subregion $A$, and analogously for $S_B$, $S_C$. Similarly, let $S_{AB}$ denote the entropy associated with $A\cup B$,
etc. The topological entanglement entropy  of $R$ is then defined~as \cite{KP}
$S_{\rm top} = S_A \nspc + \nspc S_B\nspc + \nspc S_C\nspc - \nspc S_{AB}\nspc - \nspc S_{AC}\nspc - \nspc S_{BC}\nspc + \nspc S_{ABC}.$
This linear combination has the special property that all non-universal perimeter-law contributions cancel out. Moreover,  any local deformation of the entangling
boundary does not alter the final result. 

Applying this definition to a topological field theory associated with a 2-D rational CFT, one finds \cite{KP,LW} that an
empty region of space has  $S_{\rm top}(0) = \log\bigl(1/D\bigr)  = \log\bigl( S_0^{\spc 0}\spc \bigr)$.
Here $D$ is the total quantum dimension of the medium, and related to the quantum dimension $d_a$ of  individual excitations via $D^2=  \sum_a d_a^2$. In case the region contains a quasi-particle excitation of charge $a$, the topological entanglement entropy is given by the formula \cite{fradkin}
\setcounter{equation}{2}
\bea
\label{stv}
S_{\rm top}(a) \is \log\bigl(d_a/D\bigr) = \log\bigl( S_0^{\spc a}\bigr).
\eea
Note that for a rational CFT the topological entanglement is always negative since $d_a~<~D$.
By itself, this would not make sense, as the definition of the entanglement entropy is 
a manifestly positive quantity. Once we include the non-universal contribution proportional to
the length $L$ of the entangling boundary, however, the total result is positive.

\setcounter{equation}{26}

We make the assumption 
that the relationship between the quantum dimension and the topological entanglement entropy 
remains mostly unchanged in going from rational  to non-rational CFTs. 
One important difference is that there no longer
exists an analog of the total quantum dimension $D$, and hence there is no obvious
notion of topological entanglement entropy of an empty region of space. However, there does 
exist a natural formula for the topological entanglement entropy of a black hole excitation. Applying the formula (\ref{stv}) to the hyperbolic Virasoro representation associated with a BTZ
black hole, we find
\bea
S_{\rm top}(M, J)  \is %  \log\bigl(d(\alpha_+)d(\alpha_-) \bigr) \, = \, 
\log\bigl( S^{\spc p_+}_0S^{\spc p_-}_0\bigr)\spc .
\eea
The relation between the  mass $M$ and spin $J$ and the Liouville momenta is given in 
Equation (\ref{dict}) with $\Delta_\pm$ defined in (\ref{mjrel}). Plugging in the explicit modular 
S-matrix element (\ref{smeas}) gives
\bea
\label{stops}
S_{\rm top}(M, J) \, = \, \log\Bigl(8 \sinh (2\pi b\smpc p_+\nspc ) \sinh (2\pi b^{-1}\nspc p_+\nspc)\sinh (2\pi b\smpc p_-\nspc ) \sinh (2\pi b^{-1}\nspc p_-\nspc)\Bigr).
\eea
Note that unlike the rational CFT case, the right-hand side is positive. Moreover, it grows unboundedly for large $p_\pm$. 
In the limit where $\ell M\pm J$ and $b$ are all large, it reduces to
\bea
\label{finalmatch}
S_{\rm top}(M, J) \; = \; 2\pi b\spc (p_+ + p_-) \; =
%\simeq \pi  \bigl(\sqrt{\ell \Delta_+}  + \sqrt{\ell \Delta_-}\bigr) \, \simeq
 \, \frac{2\pi r_+}{4},
\eea
which exactly matches the Bekenstein-Hawking entropy. This is our main result.

\bigskip
\bigskip

\noindent
{\large \bf 6. Higher Spin Black Hole Entropy}

\medskip

As a test of our proposal, let us consider black holes in 2+1-D higher spin gravity \cite{HS,GK}. 
Luckily, all the necessary technology is available. Our presentation will be brief.

Higher spin gravity in 2+1 dimensions is a generalization of Einstein gravity in 2+1 dimensions that includes a collection of $n-2$ higher spin fields \cite{HS}. All the fields together can be assembled into
a $SL(n, \mathbb{R}) \times SL(n,\mathbb{R})$ gauge connection $(A_+, A_-)$ with a Chern-Simons action.
The generalized space-time geometry of a higher spin black hole is characterized by two $SL(n, \mathbb{R})$ holonomies
\bea
\label{quotienttwo}
h_\pm = e^{\mbox{\small $2\pi (\lambda_+ \pm \lambda_-)/\ell$}},
\eea
which generalize the $SL(2,\mathbb{R})$ holonomies (\ref{quotient}) of the BTZ black hole. 
Here $\ell$ denotes the higher spin generalization of the AdS$_3$ radius, and $\lambda_+$ and $\lambda_-$ are diagonal elemenents of the $\mathfrak{sl}(n,\mathbb{R})$ Lie algebra. Higher spin black holes thus carry $2(n-1)$ quantum numbers, including the mass,
angular momentum and 2$(n\!-\!2)$ higher spin charges. 

Extracting an actual space-time geometry  from this general description of the
higher spin black hole turns out to be a non-trivial task. In particular, there is no gauge-invariant notion of a 2+1-D space-time metric that can be used to compute
a horizon area. As a result, there appears to be no immediately obvious  higher spin
generalization of the Bekenstein-Hawking formula. There are indeed various proposals \cite{GK, Jan}.

A simple geometric proposal for a generalized Bekenstein-Hawking formula was put forward in \cite{Jan}.
Let $e_i$ denote the simple roots of $\mathfrak{sl}(n)$ and $\langle \ , \ \rangle$ denote the Cartan Killing form.  %The fundamental weights  $\omega_i$ are vectors
%such that $\langle \omega_i, e_j\rangle = \delta_{ij}$.  
The Weyl vector is defined as $\rho = \frac 12 \sum_{e>0} e$. The higher spin generalization of the black hole entropy formula derived in \cite{Jan} is expressed in terms of the $SL(n,\mathbb{R})$ holonomies $h_{\pm}$ as
\bea
\label{shsbh}
S_{\rm HSBH}  \is \frac{ 2\pi} {4} \langle\spc \rho\smpc , \lambda_+ \rangle\, .
\eea
This elegant proposal passes some non-trivial checks \cite{Jan} and appears well-motivated.

Can one reproduce the generalized B-H  formula (\ref{shsbh}) 
by counting states in the dual CFT? This is a non-trivial task, since one needs a generalization of the
Cardy formula that keeps track of conformal dimensions and all higher spin quantum numbers. This has not been done yet.
We now give a simple derivation of (\ref{shsbh}) via the topological entanglement entropy (\ref{stop}).

2+1-D higher spin gravity is dual to 1+1-D conformal field theory with $W_n$ symmetry, the natural higher spin generalization of Virasoro symmetry. The universal CFT with $W_n$ symmetry is $\mathfrak{sl}(n,\mathbb{R})$ Toda theory
\bea
S =\frac 1 {2\pi} \int\!\! d^2\xi\, \Bigl[\spc \langle \partial\smpc \varphi, \bar\partial \varphi\rangle + R\langle Q, \varphi\rangle + \mu\sum e^{b \smpc \langle e_i ,\varphi\rangle}\Bigr]\, , \quad 
 \quad Q = 2(b + b^{-1}) \rho.
\eea
Toda theory is a non-rational CFT with central $c = n-1 + 3 \spc \langle \smpc Q\smpc , \smpc Q\smpc \rangle$. As before, states of the 2+1-D higher spin theory
with localized excitations are identified with the tensor product of left and right conformal blocks of the CFT. Black holes states correspond to vertex operators that
create macroscopic holes in the generalized space time, with holonomies (\ref{quotienttwo})
in a hyperbolic conjugacy class of $SL(n,\mathbb{R})$.
Their vertex operators $V = e^{\langle \alpha_+, \varphi_+\rangle}e^{\langle \alpha_-, \varphi_-\rangle}$ have Toda momenta $\alpha_\pm\! = \frac 1 2 Q  + i p_\pm$ and  conformal weights 
$\Delta_\pm\! = \nspc \langle \alpha_\pm\nspc , Q \nspc -\nspc \alpha_\pm\rangle\nspc =
\nspc \langle p_\pm,p_\pm \rangle + \frac 1 4 \langle Q, Q\rangle$. The semi-classical relations (\ref{semrel})  naturally generalize to
\bea
 \lambda_\pm = 4 \smpc b \spc  (p_+ \pm p_-) , \quad & & \quad
b^2 \langle \spc \rho, \rho\spc \rangle = {\ell}/{8}.
\eea

Just like their BTZ counterparts, higher spin black holes can be viewed as macroscopic quasi-particle excitations with topological interactions. We can thus compute their topological entanglement entropy
in the same way as before. The relevant modular S-matrix elements of  $\mathfrak{sl}(n,\mathbb{R})$ Toda field theory was computed in \cite{NDJ}
\bea
S^{\, p}_0 =  \Xi \,\prod_{e>0} 4 \sinh\bigl(\pi b \langle e,p\rangle\bigr) \sinh\bigl(\pi b^{-1}\langle e,p\rangle\bigr)
\eea
with $\Xi$ some irrelevant constant. % =  i^{n-1} \sqrt{{\det C}}/|{\cal W}|$ with $C$ the Cartan matrix and $|{\cal W}$ the order of the Weyl group. 
Using the formula $S_{\rm top} = \log\bigl(S^{\spc p_+}_0 S^{\spc p_-}_0\bigr)$ and taking
the semi-classical limit, we reproduce the result (\ref{shsbh})
\bea
S_{\rm top} \, = \,  2\pi b\spc \bigl(  \langle\smpc  \rho , \smpc p_+\rangle + \langle \rho, p_-\spc \rangle\bigr) \, = \, \frac{2\pi}{4}\spc \langle \rho, \lambda_+\rangle.
\eea

\bigskip
\bigskip

\noindent
{\bf \large 7. Concluding Remarks}

\medskip

We have put forward a new interpretation of 2+1-D quantum gravity as the effective field theory that describes the long range properties of a highly entangled ground state.
In line with this interpretation, we have computed the topological entanglement entropy of 
a BTZ black hole. Our computation  does not make use of the Bekenstein-Hawking, Ryu-Takayanagi, 
or Cardy formulas. It is a new and independent derivation,  yet yields a leading-order result that matches all three.  
 Our result also raises a number of questions. We briefly comment on some of them.

\medskip
\medskip

\noindent
{\it  Does pure 2+1-D quantum gravity exist?  What is its role?}

\smallskip

Via the identification with the space of left and right conformal blocks of 2-D Liouville theory, we have given a well-defined description of the Hilbert space of 2+1-D quantum gravity. Does this mean that pure 2+1-D 
quantum gravity exists as a UV complete theory? The answer is ``No" \cite{MW}. The spectrum of Virasoro representations 
is continuous, and thus the level density of states  of Liouville theory and pure 2+1-D gravity is strictly infinite. 
This is an unphysical situation. 
To get a well-behaved physical system, one needs  to supply a specific holographic dual in the form of some unitary 2-D CFT. This CFT prescribes the allowed discrete spectrum of conformal dimensions, with a finite level density. In this note, we implicitly assumed that this CFT is maximally non-rational, i.e. that it does not have 
any other symmetries than conformal invariance. In this idealized case, once the spectrum of excitations
is prescribed, 2+1-D gravity gives an accurate description of their long range interactions and assigns the correct quantum dimension to the black hole states.

\medskip
\medskip

\noindent
{\it What does the topological entanglement entropy count?}

\medskip

This is the most important question. It is natural to interpret $S_{\rm top}$ as 
the universal contribution to the entanglement across the black hole horizon. The fact that it saturates the B-H bound
is consistent with the idea \cite{van} that entanglement is responsible for the 
continuity of space across the horizon. However,
this interpretation immediately raises an important puzzle, closely related to the firewall paradox \cite{AMPS}.

According to the usual AdS/CFT dictionary, any typical CFT state with large enough energy 
describes a black hole in the bulk. The level density of the CFT indeed matches the microscopic B-H entropy.
However, to write a state with entanglement entropy proportional to $S_{\rm BH}$, one 
needs to include {\it two} Hilbert space sectors each with entropy at least equal to $S_{\rm BH}$.
The CFT seems to provide only one of these sectors. So where is the other sector?

  Liouville vertex operators with momenta
$\alpha = \frac 1 2 Q + i p$ in fact create macroscopic holes in space, as indicated in Fig.~3.
Based on the similarity with Fig.~1, it is tempting to identify both sides of the hyperboloid in
Fig.~3 as the two sides of the eternal black hole solution. According to this interpretation, it seems that
by acting with the vertex operator, one has created a completely new asymptotic region with its own
holographic CFT dual. This could be where the other sector resides.
But how would one create such a second asymptotic region via gravitational collapse,
i.e. by acting with operators on the vacuum of one single CFT?  This is one version of the firewall question.

\medskip
\medskip

\noindent
{\it Is there a firewall or fuzzball? Is $S_{\rm BH}$ a boundary entropy?}

\medskip

In our view, if our proposal that the  entanglement entropy of BTZ black holes saturates the
B-H bound is correct, then there is no firewall. The state looks like an eternal black hole 
that realizes the balanced holography postulate put forward in \cite{VV}.
The entanglement across the horizon is then sufficient to safeguard the continuity of space \cite{van}.

There is, however, another possible interpretation\footnote{We thank Nick Warner for emphasizing this possible alternative interpretation.} of our formula $S_{\rm BH} = \log S^{\, a}_0$ in terms 
of the Affleck-Ludwig boundary entropy \cite{AL}. Suppose that, instead of the hyperbolic solution 
of Fig.~3,  we place a reflecting boundary at the location of the black hole horizon. A natural conformal boundary for Liouville CFT
 is the ZZ-boundary state $| \spc  ZZ\spc \rangle$ \cite{ZZ}. Its overlap
 with the Ishibashi state $|\! | \spc p \spc \rangle \!\nspc\rangle$, the eigenstates with given Liouville momentum $\alpha  = \frac 1 2 Q + i p$, satisfies
\bea
 \bigl| \Psi_{\rm ZZ}(p) \bigr|^2 = 
S_0^{\spc p} , \quad & & \quad \Psi_{\rm ZZ} (p) =  \langle\!\!\smpc\langle\spc \raisebox{1pt}{$p$}\, |\!\nspc\smpc | \spc{ZZ} \spc \rangle.
\eea
This implies that the boundary entropy of the ZZ state in the sector with momentum $p$ is equal to 
$\log(S_0^p)$. Moreover, if we identify the topological entanglement entropy with the Bekenstein-Hawking
entropy of the BTZ black hole, we obtain the very suggestive relation
\bea
Z_{\rm BH} =\bigl| \Psi_{\rm ZZ}(p_+,p_-)\bigr|^2
\eea
with $Z_{\rm BH} = e^{S_{\rm BH}}$.
Could it be that, instead of topological entanglement entropy, our formula is counting the boundary
entropy of a reflecting boundary at the horizon? Or are both interpretations correct? 
Either way, we believe that finding the answer to these questions will shed important 
new light on the nature of the interior geometry of black holes.

\vspace{5mm}

\begin{center}{\large \bf Acknowledgement}
\end{center}

We thank Daniel Harlow, Tatsuma Nishioka, Per Kraus, Erik Verlinde, Nick Warner and Masahito Yamazaki for helpful discussions.
The work is supported by NSF grant PHY-1314198.

\bigskip
\bigskip

\addtolength{\baselineskip}{-.3mm}


\begin{thebibliography}{99}
\bibitem{BH} S. Hawking, %Particle Creation by Black Holes, 
Commun.Math.Phys. 43 (1975) 199;   J.~D.~Bekenstein, %``Black holes and entropy,'' 
 Phys.\ Rev.\ D {\bf 7}, 2333 (1973).
 \bibitem{SV} 
  A.~Strominger and C.~Vafa,
  %``Microscopic origin of the Bekenstein-Hawking entropy,''
  Phys.\ Lett.\ B {\bf 379}, 99 (1996)
  [hep-th/9601029].
  %%CITATION = HEP-TH/9601029;%%
  %1696 citations counted in INSPIRE as of 08 Aug 2013
  %\cite{Bombelli:1986rw}
\bibitem{GEE}
  L.~Bombelli, R.~K.~Koul, J.~Lee and R.~D.~Sorkin,
  %``A Quantum Source of Entropy for Black Holes,''
  Phys.\ Rev.\ D {\bf 34} (1986) 373;
  %%CITATION = PHRVA,D34,373;%%
  %488 citations counted in INSPIRE as of 09 Aug 2013%\cite{Srednicki:1993im}
  M.~Srednicki,
  %``Entropy and area,''
  Phys.\ Rev.\ Lett.\  {\bf 71}, 666 (1993)
  [hep-th/9303048];
  %%CITATION = HEP-TH/9303048;%%
  %489 citations counted in INSPIRE as of 09 Aug 2013
  %\cite{Callan:1994py}
  C.~G.~Callan, Jr. and F.~Wilczek,
  %``On geometric entropy,''
  Phys.\ Lett.\ B {\bf 333}, 55 (1994)
  [hep-th/9401072].
  %%CITATION = HEP-TH/9401072;%%
  %287 citations counted in INSPIRE as of 09 Aug 2013
  \bibitem{RT} 
  S.~Ryu and T.~Takayanagi,
  %``Holographic derivation of entanglement entropy from AdS/CFT,''
  Phys.\ Rev.\ Lett.\  {\bf 96}, 181602 (2006)
  [hep-th/0603001].
\bibitem{van}  M. Van Raamsdonk, Comments on quantum gravity and entanglement, 0907.2939; 
 J.~Maldacena and L.~Susskind,
  %``Cool horizons for entangled black holes,''
  arXiv:1306.0533 [hep-th].
  %%CITATION = ARXIV:1306.0533;%%
  %16 citations counted in INSPIRE as of 07 Aug 2013
\bibitem{AMPS}
  %A. Almheiri, D. Marolf, J. Polchinski and J. Sully, Black Holes: Complementarity or Firewalls?, 1207.3123
  A. Almheiri, D. Marolf, J. Polchinski and J. Sully, 
  %"Black Holes: Complementarity or Firewalls?," 
  JHEP {\bf 1302}, 062 (2013) [arXiv:1207.3123 [hep-th]];  
  %S. L. Braunstein, 
  %"Black hole entropy as entropy of entanglement, or it's curtains for the equivalence principle," published as 
  cf. S. L. Braunstein, S. Pirandola and K. Zyczkowski, 
  %"Better Late than Never: Information Retrieval from Black Holes," 
  Phys.\ Rev.\ Lett.\ {\bf 110}, 101301 (2013) [arXiv:0907.1190v1 [quant-ph]] for a similar prediction from different assumptions.
  \bibitem{WG}
  E.~Witten,
  %``(2+1)-Dimensional Gravity as an Exactly Soluble System,''
  Nucl.\ Phys.\ B {\bf 311}, 46 (1988).
  %%CITATION = NUPHA,B311,46;%%
  %1333 citations counted in INSPIRE as of 07 Aug 2013
  \bibitem{WCS}%\cite{Witten:1988hf} 
  E.~Witten,
  %``Quantum Field Theory and the Jones Polynomial,''
  Commun.\ Math.\ Phys.\  {\bf 121}, 351 (1989).
  %%CITATION = CMPHA,121,351;%%
\bibitem{KP} 
  A.~Kitaev and J.~Preskill,
  %``Topological entanglement entropy,''
  Phys.\ Rev.\ Lett.\  {\bf 96}, 110404 (2006)
  [hep-th/0510092].
  %%CITATION = HEP-TH/0510092;%%
  %165 citations counted in INSPIRE as of 07 Aug 2013
%\cite{Levin:2006zz}
\bibitem{LW} 
  M.~Levin and X.~-G.~Wen,
  %``Detecting Topological Order in a Ground State Wave Function,''
  Phys.\ Rev.\ Lett.\  {\bf 96}, 110405 (2006).
  %%CITATION = PRLTA,96,110405;%%
  %139 citations counted in INSPIRE as of 07 Aug 2013
  %\cite{Banados:1992gq}%\cite{Banados:1992wn}
%\cite{Dong:2008ft}
\bibitem{nayak}
  C.~Nayak, S.~H.~Simon, A.~Stern, M.~Freedman, and S.~Das Sarma,
  %``Non-Abelian anyons and topological quantum computation"
  Rev.\ Mod.\ Phys.\ {\bf 80}, 1083 (2008).
\bibitem{fradkin} 
  S.~Dong, E.~Fradkin, R.~G.~Leigh and S.~Nowling,
  %``Topological Entanglement Entropy in Chern-Simons Theories and Quantum Hall Fluids,''
  JHEP {\bf 0805}, 016 (2008)
  [arXiv:0802.3231 [hep-th]].
  %%CITATION = ARXIV:0802.3231;%%
  %16 citations counted in INSPIRE as of 17 Sep 2013
\bibitem{BTZ} 
  M.~Banados, C.~Teitelboim and J.~Zanelli,
  %``The Black hole in three-dimensional space-time,''
  Phys.\ Rev.\ Lett.\  {\bf 69}, 1849 (1992)
  [hep-th/9204099];
  %%CITATION = HEP-TH/9204099;%%
  %1559 citations counted in INSPIRE as of 07 Aug 2013
  M.~Banados, M.~Henneaux, C.~Teitelboim and J.~Zanelli,
  %``Geometry of the (2+1) black hole,''
  Phys.\ Rev.\ D {\bf 48}, 1506 (1993)
  [gr-qc/9302012].
  %%CITATION = GR-QC/9302012;%%
  %970 citations counted in INSPIRE as of 07 Aug 2013 
  \bibitem{VV} 
  E.~Verlinde and H.~Verlinde,
  %``Passing through the Firewall,''
  arXiv:1306.0515 [hep-th].
  %%CITATION = ARXIV:1306.0515;%%
  %5 citations counted in INSPIRE as of 07 Aug 2013
  %%CITATION = HEP-TH/0310234;%%
  %17 citations counted in INSPIRE as of 07 Aug 2013
  %1944 citations counted in INSPIRE as of 07 Aug 2013
  \bibitem{HV}
  H.~L.~Verlinde,
  %``Conformal Field Theory, 2-d Quantum Gravity And Quantization Of Teichmuller Space,''
  Nucl.\ Phys.\ B {\bf 337}, 652 (1990).
  %%CITATION = NUPHA,B337,652;%%
  %89 citations counted in INSPIRE as of 07 Aug 2013
  \bibitem{carlip}  
  S.~Carlip,
  %``Conformal field theory, (2+1)-dimensional gravity, and the BTZ black hole,''
  Class.\ Quant.\ Grav.\  {\bf 22}, R85 (2005)
  [gr-qc/0503022];
  %%CITATION = GR-QC/0503022;%%
  %103 citations counted in INSPIRE as of 09 Aug 2013
  Y.~-j.~Chen,
  %``Quantum Liouville theory and BTZ black hole entropy,''
  Class.\ Quant.\ Grav.\  {\bf 21}, 1153 (2004)
  [hep-th/0310234].
  \bibitem{ZZ} 
  A.~B.~Zamolodchikov and A.~B.~Zamolodchikov,
  %``Liouville field theory on a pseudosphere,''
  hep-th/0101152.
  %%CITATION = HEP-TH/0101152;%%
  %207 citations counted in INSPIRE as of 07 Aug 2013
  \bibitem{TT1} 
  J.~Teschner,
  %``Liouville theory revisited,''
  Class.\ Quant.\ Grav.\  {\bf 18}, R153 (2001)
  [hep-th/0104158].
  %%CITATION = HEP-TH/0104158;%%
  %179 citations counted in INSPIRE as of 07 Aug 2013
  %\cite{Teschner:2008qh}
  \bibitem{TT2} 
  J.~Teschner,
  %``On the relation between quantum Liouville theory and the quantized Teichmuller spaces,''
  Int.\ J.\ Mod.\ Phys.\ A {\bf 19S2}, 459 (2004)
  [hep-th/0303149];  
  %``Nonrational conformal field theory,''
  arXiv:0803.0919 [hep-th].
\bibitem{BRH} 
  J.~D.~Brown and M.~Henneaux,
  %``Central Charges in the Canonical Realization of Asymptotic Symmetries: An Example from Three-Dimensional Gravity,''
  Commun.\ Math.\ Phys.\  {\bf 104}, 207 (1986).
  %%CITATION = CMPHA,104,207;%%
  %900 citations counted in INSPIRE as of 08 Aug 2013
  \bibitem{Nati}
  N.~Seiberg,
  %``Notes on quantum Liouville theory and quantum gravity,''
  Prog.\ Theor.\ Phys.\ Suppl.\  {\bf 102}, 319 (1990).
  %%CITATION = PTPSA,102,319;%%
  %281 citations counted in INSPIRE as of 10 Aug 2013
  %\bibitem{chen}
  %Y.~-j.~Chen,
  %%``Quantum Liouville theory and BTZ black hole entropy,''
  %Class.\ Quant.\ Grav.\  {\bf 21}, 1153 (2004)
  %[hep-th/0310234].
  %%CITATION = HEP-TH/0310234;%%
  %18 citations counted in INSPIRE as of 10 Sep 2013
  \bibitem{EV} E.~P.~Verlinde,
  %``Fusion Rules and Modular Transformations in 2D Conformal Field Theory,''
  Nucl.\ Phys.\ B {\bf 300}, 360 (1988);  J.~L.~Cardy,
  %``Boundary Conditions, Fusion Rules and the Verlinde Formula,''
  Nucl.\ Phys.\ B {\bf 324}, 581 (1989);  R.~Dijkgraaf and E.~P.~Verlinde,
  %``Modular Invariance And The Fusion Algebra,''
  Nucl.\ Phys.\ Proc.\ Suppl.\  {\bf 5B}, 87 (1988).
  %%CITATION = NUPHZ,5B,87;%%
  %32 citations counted in INSPIRE as of 10 Aug 2013
  %%CITATION = NUPHA,B324,581;%%
  %649 citations counted in INSPIRE as of 07 Aug 2013
  %%CITATION = NUPHA,B300,360;%%
  %671 citations counted in INSPIRE as of 07 Aug 2013
  %\cite{Affleck:1991tk}
\bibitem{HS} 
  M.~R.~Gaberdiel and R.~Gopakumar,
  %``Minimal Model Holography,''
  J.\ Phys.\ A {\bf 46}, 214002 (2013)
  [arXiv:1207.6697 [hep-th]].
  %%CITATION = ARXIV:1207.6697;%%
  %54 citations counted in INSPIRE as of 10 Aug 2013
\bibitem{GK} 
  M.~Ammon, M.~Gutperle, P.~Kraus and E.~Perlmutter,
  %``Black holes in three dimensional higher spin gravity: A review,''
  J.\ Phys.\ A {\bf 46}, 214001 (2013)
  [arXiv:1208.5182 [hep-th]].
  %%CITATION = ARXIV:1208.5182;%%
  %38 citations counted in INSPIRE as of 10 Aug 2013
  \bibitem{Jan} 
  J.~de Boer and J.~I.~Jottar,
  %``Thermodynamics of Higher Spin Black Holes in AdS$_{3}$,''
  arXiv:1302.0816 [hep-th], arXiv:1306.4347 [hep-th];
  %%CITATION = ARXIV:1306.4347;%%
  %3 citations counted in INSPIRE as of 10 Aug 2013
  %%CITATION = ARXIV:1302.0816;%%
  %11 citations counted in INSPIRE as of 10 Aug 2013
  A.~Perez, D.~Tempo and R.~Troncoso,
  %``Higher spin gravity in 3D: black holes, global charges and thermodynamics,''
  arXiv:1207.2844 [hep-th].
  %%CITATION = ARXIV:1207.2844;%%
  %22 citations counted in INSPIRE as of 10 Sep 2013
  \bibitem{NDJ} 
  N.~Drukker, D.~Gaiotto and J.~Gomis,
  %``The Virtue of Defects in 4D Gauge Theories and 2D CFTs,''
  JHEP {\bf 1106}, 025 (2011)
  [arXiv:1003.1112 [hep-th]].
  %%CITATION = ARXIV:1003.1112;%%
  %78 citations counted in INSPIRE as of 10 Aug 2013
  %\cite{Maloney:2007ud}
\bibitem{MW} 
  A.~Maloney and E.~Witten,
  %``Quantum Gravity Partition Functions in Three Dimensions,''
  JHEP {\bf 1002}, 029 (2010)
  [arXiv:0712.0155 [hep-th]].
  %%CITATION = ARXIV:0712.0155;%%
  %105 citations counted in INSPIRE as of 10 Aug 2013
\bibitem{AL} 
  I.~Affleck and A.~W.~W.~Ludwig,
  %``Universal noninteger 'ground state degeneracy' in critical quantum systems,''
  Phys.\ Rev.\ Lett.\  {\bf 67}, 161 (1991).
  %%CITATION = PRLTA,67,161;%%
  %255 citations counted in INSPIRE as of 09 Aug 2013
  %\cite{Gaberdiel:2012uj}
\end{thebibliography}
\end{document}